\documentclass[sigconf]{acmart}
\renewcommand\footnotetextcopyrightpermission[1]{}
\usepackage{graphicx}
\setkeys{Gin}{draft=False}

\usepackage{amsmath}

\usepackage{amssymb}
\usepackage{booktabs}
\usepackage{multirow}
\usepackage{array}
\usepackage{url}
\usepackage{threeparttable}
\usepackage{enumitem}
\usepackage{microtype}
\usepackage{balance}  
\usepackage{placeins} 

\usepackage{tabularx}

\usepackage{siunitx}
\usepackage{pgfplots}
\pgfplotsset{compat=1.18}

\usepackage{siunitx}
\AtBeginDocument{%
  }

\setcopyright{acmlicensed}
\copyrightyear{2025}
\acmYear{2025}
\acmDOI{XXXXXXX.XXXXXXX}

\acmConference[XX]{XX}{XX}{XX}

\acmISBN{978-1-4503-XXXX-X/2018/06}

\begin{document}

\title{Beyond Patch Aggregation: 3-Pass Pyramid Indexing for Vision-Enhanced Document Retrieval}

\author{Anup Roy}
\email{anup.roy@inceptionai.ai}
\affiliation{%
  \institution{Inception AI}
  \city{Abu Dhabi}
  \country{UAE}
}

\author{Rishabh Gyanendra Upadhyay}
\email{rishabh.upadhyay@inceptionai.ai}
\affiliation{%
  \institution{Inception AI}
  \city{Abu Dhabi}
  \country{UAE}
}

\author{Animesh Rameshbhai Panara}
\email{animesh.panara@inceptionai.ai}
\affiliation{%
  \institution{Inception AI}
  \city{Abu Dhabi}
  \country{UAE}
}

\author{Robin Mills}
\email{robin.mills@inceptionai.ai}
\affiliation{%
  \institution{Inception AI}
  \city{Abu Dhabi}
  \country{UAE}
}

\author{Aidan Philip Millar}
\email{amillar@mubadala.ae}
\affiliation{%
  \institution{Mubadala}
  \city{Abu Dhabi}
  \country{UAE}
}

\renewcommand{\shortauthors}{Roy et al.}

\begin{abstract}

Document-centric RAG pipelines typically begin with OCR, followed by brittle, engineering-heavy heuristics for chunking, table parsing, and layout reconstruction. These text-first workflows are costly to maintain, sensitive to small layout shifts, and discard the visuo-spatial cues that frequently contain the answer. Vision-first retrieval has recently emerged as a compelling alternative: by operating directly on page images, systems such as ColPali and ColQwen preserve spatial structure and reduce pipeline complexity while achieving strong benchmark performance. However, these late-interaction models tightly couple retrieval to a specific vision backbone and require storing hundreds of patch embeddings per page, creating substantial memory overhead and complicating large-scale deployment. We introduce \textbf{VisionRAG}, a multimodal retrieval system that is both \emph{OCR-free} and \emph{model-agnostic}. VisionRAG indexes documents directly as images, preserving layout, table structure, and spatial cues, and constructs semantic vectors without committing to a specific extraction. Our \textbf{three-pass pyramid indexing} framework create  semantic vectors using global page summaries, section headers, visual hotspots, and fact-level cues. These summaries serve as lightweight retrieval surrogates: at query time, VisionRAG retrieves the most relevant pages using the pyramid index, then forwards the \emph{raw page image} (encoded as base64) to a multimodal LLM for final question answering. During retrieval, \textbf{reciprocal rank fusion} integrates representations across the pyramid, yielding robust ranking across heterogeneous visual and textual content. VisionRAG maintains just \textbf{17-27 vectors per page}, matching the efficiency of patch-based approaches while remaining adaptable to different multimodal encoders. On financial document benchmarks, VisionRAG achieves \textbf{0.8051 accuracy@10} on \textbf{FinanceBench} and \textbf{0.9629 Recall@100} on \textbf{TAT-DQA}, demonstrating strong coverage of answer-bearing content in complex, visually rich documents. These results suggest that OCR-free, summary-guided multimodal retrieval provides a practical and scalable alternative to traditional text-extraction pipelines.
\end{abstract}



\maketitle

\section{Introduction}

Retrieval-Augmented Generation (RAG) has improved factual grounding in large language models by enabling access to external knowledge sources \cite{rag2020}. However, in enterprise and financial domains, critical information is embedded in \emph{visually rich} PDFs containing complex tables, multi-column layouts, section hierarchies, and spatial cues. OCR-based pipelines flatten these structures into plain text, discarding layout boundaries, table geometry, and reading order-leading to degraded retrieval recall and weaker downstream answer quality.

These limitations are amplified in document-intensive settings such as financial filings, where hundreds of densely formatted pages contain key facts within table cells, visually emphasized regions, or multi-column spans that OCR systems often fragment or misinterpret. As a result, text-only representations fail to capture the multimodal signals necessary for accurate indexing and retrieval.

\begin{figure*}[!t]
\centering
\includegraphics[width=0.70\textwidth]{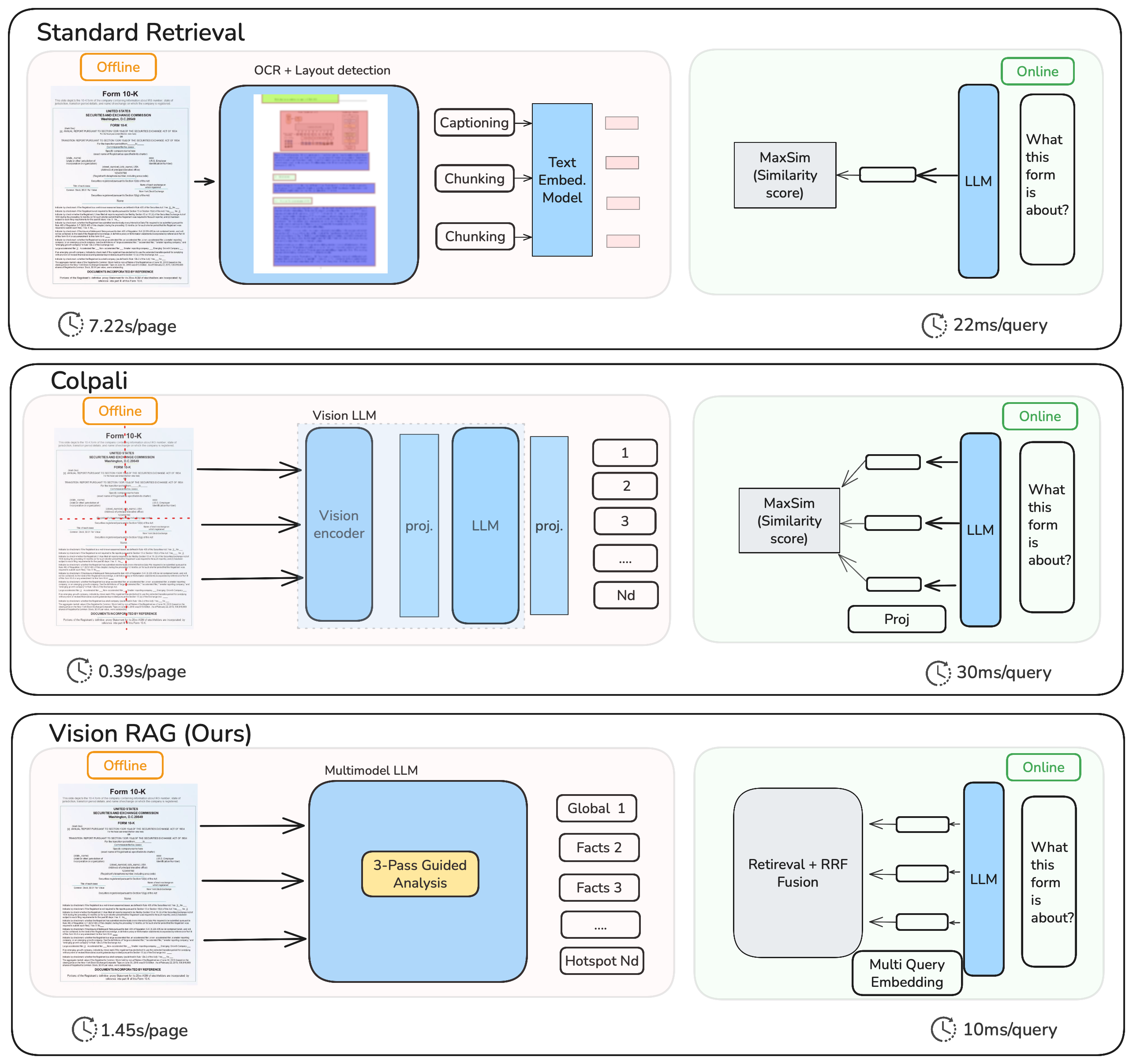}
\caption{Evolution of document retrieval approaches. (Top) OCR-based RAG flattens visual structure, losing layout and table context. (Middle) ColPali adds vision awareness via dense patch embeddings (\textasciitilde1{,}024 vectors/page) but at high cost. (Bottom) VisionRAG introduces pyramid indexing with semantic fusion across page, section, and fact levels, achieving similar accuracy with only 12--17 vectors per page and no OCR dependency.}

\label{fig:rag_evolution}
\end{figure*}

Recent vision-aware systems address these issues by processing document pages directly as images. Approaches such as ColPali generate dense patch-level embeddings to support image-to-text matching. While effective, they impose substantial computational cost: ColPali produces multi-dimension embeddings per page, and even aggressively pooled variants still require $\sim$341 vectors-far exceeding what is feasible for large-scale indexing and low-latency retrieval. Figure~\ref{fig:rag_evolution} summarizes this evolution from OCR-based RAG to dense vision retrieval and our proposed approach.

These computational constraints pose a challenge in enterprise environments, where repositories may contain millions of pages and latency, memory, and hardware budgets are tightly restricted. VisionRAG is designed specifically with these constraints in mind. By relying on compact semantic representations (17-27 vectors per page), minimizing GPU dependence, and remaining compatible with standard vector search engines (e.g., FAISS, Elasticsearch ANN, Milvus), VisionRAG offers a more practical accuracy-latency-cost profile for production deployment.

To address the shortcomings of both OCR-based and patch-based vision retrieval, we introduce \textbf{VisionRAG}, an OCR-free, multimodal retrieval system built around three principles:

\begin{enumerate}
    \item \textbf{Page-as-image semantic analysis.} Each page is processed as a high-resolution image by a vision-language model (VLM), producing complementary textual signals including page summaries, section descriptions, fact-level cues, and visual hotspot interpretations.
    \item \textbf{Pyramid indexing.} These signals form four lightweight indices-page, section, fact, and hotspot-each optimized for different query granularities and information needs, avoiding dense patch embeddings entirely.
    \item \textbf{Explicit semantic fusion.} VisionRAG performs content-level fusion prior to embedding and ranking-level fusion via reciprocal rank fusion (RRF), yielding interpretable retrieval behavior and a compact footprint (17-27 vectors per page vs.\ 1{,}024 for ColPali).
\end{enumerate}

\begin{figure}[!t]
  \centering
  \includegraphics[width=0.48\textwidth]{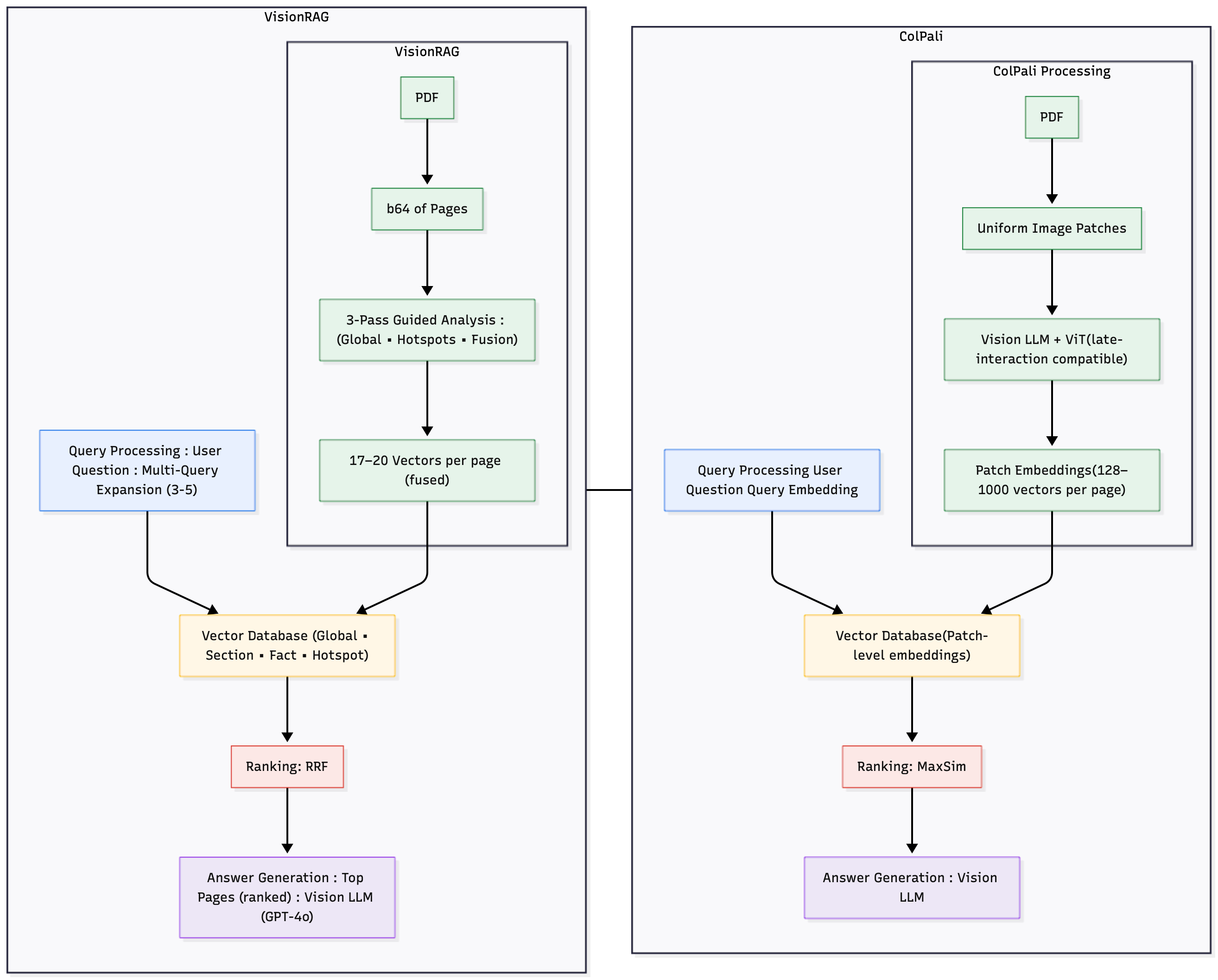}
  \captionsetup{font=small}
\caption{High-level comparison. \textbf{Right:} ColPali encodes dense patch embeddings with late interaction. \textbf{Leftt:} VisionRAG builds compact multi-level indices and fuses results via Reciprocal Rank Fusion (RRF).}
\label{fig:colpali_comparison}
\end{figure}

At query time, VisionRAG retrieves the most relevant pages using the pyramid index and forwards the \emph{raw page image} (encoded as base64) to a VLM for final question answering (Figure~\ref{fig:colpali_comparison}). This decouples retrieval from model-specific patch embeddings and avoids the quadratic interaction costs of deep late-interaction systems. By achieving a favorable accuracy-latency-cost Pareto frontier, VisionRAG provides efficient indexing, fast ANN search, and robust RRF fusion while maintaining strong coverage at $K{=}100$ and high end-to-end QA performance. This makes VisionRAG a practical and scalable alternative to both OCR-based RAG and heavy patch-based vision retrieval models.

\subsection{Contributions}

Our key contributions are as follows:

\begin{itemize}
    \item \textbf{OCR-free vision processing.} We eliminate OCR-based text extraction entirely, processing documents directly in their visual form while preserving layout, spatial structure, and table geometry.
    \item \textbf{Model-agnostic multimodal design.} VisionRAG supports diverse document types and VLM architectures without modality-specific preprocessing, enabling flexible integration into heterogeneous pipelines.
    \item \textbf{Pyramid indexing with explicit fusion.} We introduce a lightweight, interpretable alternative to dense patch-level fusion, combining page-, section-, fact-, and hotspot-level signals through content-level and ranking-level fusion.
    \item \textbf{Comprehensive evaluation.} We provide detailed experiments, ablations, and cross-model comparisons on challenging financial benchmarks, demonstrating strong retrieval and QA performance.
    \item \textbf{Production-ready efficiency.} We analyze memory footprint, indexing overhead, and retrieval latency, showing that VisionRAG’s 17-27 vectors per page yield a substantially more practical accuracy-latency-cost tradeoff than patch-based vision retrieval approaches.
\end{itemize}

\section{Related Work}
\label{sec:related}
This section situates our work within four relevant research areas: retrieval-augmented generation, visually rich document understanding, late-interaction retrieval, and query expansion. We highlight the strengths and limitations of each line of work and show how VisionRAG fills the gap between vision-aware retrieval accuracy and production-feasible efficiency.

\subsection{Retrieval-Augmented Generation}
Retrieval-Augmented Generation couples transformers with non-parametric retrieval modules, allowing models to access information without training or finetuning~\cite{rag2020}. Architectures typically involve query reformulation, retrieval, and conditional generation. Advances include iterative retrieval~\cite{iter_retro}, multi-hop reasoning, and adaptive retrieval policies that determine when additional evidence is required. Fusion strategies such as reciprocal rank fusion (RRF) provide robust aggregation of ranked lists from heterogeneous retrievers~\cite{rrf}. However, most RAG systems assume text-only corpora and rely on OCR, limiting their effectiveness on visually rich documents where layout and spatial cues carry critical semantic information.

\subsection{Visually Rich Document Understanding}
Layout-aware models such as LayoutLM~\cite{xu2020layoutlm} integrate text, spatial layout, and visual features to improve form understanding, receipt parsing, and classification tasks. Multimodal document QA benchmarks-including DocVQA~\cite{mathew2021docvqa}, InfographicsVQA~\cite{mathew2022infographicsvqa}, and TAT-DQA~\cite{zayats2021tatdqa}, demonstrate the need to reason jointly over text, tables, and visual structure. These approaches highlight the importance of multimodal understanding, but they rely heavily on OCR-derived text and do not directly address retrieval efficiency or large-scale indexing of page images.

\subsection{Late Interaction Retrieval}
Late interaction models such as ColBERT~\cite{colbert} encode query and document tokens independently and compute relevance via MaxSim, enabling fine-grained matching while preserving precomputed document representations. ColPali extends this paradigm to vision-language models by encoding page images into grids of patch vectors~\cite{colpali_arxiv,colpali_iclr,weaviate_lateint}. Although this avoids OCR and captures visual layout, it increases memory and computational demands due to the large number of multi-dimension embeddings. Pooling~\cite{elastic_pool,py_colpali_engine}, quantization, and specialized indexing~\cite{vespa_colpali,activeloop_colpali,lanceblog} mitigate but do not eliminate this overhead. As a result, late interaction methods remain expensive for production-scale document retrieval.

\subsection{Query Expansion and Reformulation}
Query expansion techniques improve coverage and recall by augmenting queries with additional signals. Classical pseudo-relevance feedback extracts expansion terms from initially retrieved documents, while modern neural methods generate paraphrases, salient keywords, or hypothetical relevant passages~\cite{hyde}. These methods reduce vocabulary mismatch but assume reliable text representations, making them less effective when OCR noise or complex layouts distort document content.

Overall, existing work demonstrates strong advances in multimodal understanding and late-interaction retrieval but leaves a gap between vision-aware accuracy and production-feasible efficiency. VisionRAG addresses this gap by using compact semantic representations derived from page images, enabling multimodal retrieval without dense patch embeddings or OCR dependencies.

\section{System Architectures}
\label{sec:architectures}

We compare the architectures of ColPali, an implicit fusion model based on late interaction and VisionRAG, which uses explicit fusion through a lightweight pyramid index. The two approaches differ fundamentally in how they represent document pages, how they fuse multimodal signals, and the computational tradeoffs they introduce.

\subsection{ColPali: Implicit Context Fusion}

ColPali processes each document page with a vision-language backbone (e.g., PaliGemma), producing a dense grid of contextualized patch embeddings. For a page of size $H \times W$, the encoder outputs a grid of $P_h \times P_w$ patches, where each patch $p_u$ has embedding $\mathbf{p}_u \in \mathbb{R}^d$. A standard configuration uses a $32 \times 32$ grid, yielding $|P| = 1{,}024$ patch vectors per page~\cite{drchris_colpali}.

A query $q$ is tokenized and encoded into embeddings $\{\mathbf{q}_1, \ldots, \mathbf{q}_{|Q|}\}$ with $\mathbf{q}_t \in \mathbb{R}^d$. Relevance is computed via MaxSim:

\begin{equation}
\text{score}_{\text{ColPali}}(q, p) = \sum_{t=1}^{|Q|} \max_{u=1}^{|P|} \langle \mathbf{q}_t, \mathbf{p}_u \rangle,
\label{eq:colpali}
\end{equation}

which fuses visual and semantic signals through the encoder's cross-attention and the late-interaction matching stage~\cite{weaviate_lateint,colpali_arxiv}.

\subsubsection{Complexity and Memory}

MaxSim scoring has complexity $O(|Q| \times |P| \times d)$. With typical dimensions ($|Q| \approx 20$, $|P| = 1{,}024$, $d = 128$), each query-page comparison requires approximately 2.6M multiply-accumulate operations (MACs). At scale, approximate MaxSim, hierarchical pruning, or GPU acceleration is required~\cite{vespa_colpali}.

Storage grows linearly with patch count. For $N$ pages, memory usage is:

\[
N \times |P| \times d \times 2 \text{ bytes},
\]

assuming float16. At one million pages (1,024 patches, $d = 128$), raw storage is roughly 262\,GB before indexing overhead.

\subsubsection{Patch Reduction}

Patch pooling can reduce computational cost with modest accuracy loss. A pooling factor of 3 compresses a $32 \times 32$ grid to roughly $11 \times 11$ (121 patches after padding). Practical implementations typically retain $\sim$341 active patches while preserving $\sim$97.8\% of the original retrieval quality, reducing storage and compute by roughly 66.7\%~\cite{elastic_pool,py_colpali_engine}.

\subsection{VisionRAG: Explicit Semantic Fusion}

VisionRAG adopts an explicit fusion strategy: it first extracts semantic textual artifacts from page images and then performs rank-based fusion across multiple lightweight indices.

\subsubsection{Page Analysis and Semantic Extraction}

For each page $p$ of document $d$, a vision-language model (e.g., GPT-4o) produces four complementary artifact types:
\begin{enumerate}
    \item \textbf{Page summary} ($\text{sum}_{d,p}$): 6-10 sentences describing key topics and claims.
    \item \textbf{Section headers} ($\text{sec}_{d,p} = \{h_s\}$): hierarchical headings, captions, and figure titles.
    \item \textbf{Key facts} ($\text{fact}_{d,p} = \{f_i\}$): atomic factual units such as numbers, entities, and short statements.
    \item \textbf{Visual hotspots} ($\text{hot}_{d,p} = \{s_j\}$): concise descriptions of salient regions (chart peaks, table headers, highlighted values).
\end{enumerate}

These artifacts provide different lenses on the page: summaries offer global context; headers give structural cues; facts supply high-precision atomic information; and hotspots capture visually emphasized or tabular content not easily represented through text alone.

\subsubsection{Pyramid Index Construction}

VisionRAG builds four indices
\[
\mathcal{I} = \{\text{fused page},\ \text{section},\ \text{fact},\ \text{hotspot}\},
\]
where each artifact type yields a separate retrieval pathway:

\begin{enumerate}
    \item \textbf{Fused Page index}: one vector per page, combining global summary and hotspot information.
    \item \textbf{Section index}: one vector per header $h_s$:
    \[
    \mathbf{v}_{\text{sec}}(d,p,s) = \phi(h_s).
    \]
    \item \textbf{Fact index}: one vector per factual unit $f_i$:
    \[
    \mathbf{v}_{\text{fact}}(d,p,i) = \phi(f_i).
    \]
    \item \textbf{Hotspot index}: one vector per hotspot $s_j$:
    \[
    \mathbf{v}_{\text{hot}}(d,p,j) = \phi(s_j).
    \]
\end{enumerate}

A typical page produces $S \approx 2$-4 sections, $F \approx 5$-8 facts, and $H \approx 2$-4 hotspots, leading to
\[
B = 1 + S + F + H \approx 11\text{-}17
\]
vectors per page (median $\sim$12), substantially lower than patch-based approaches.

\subsubsection{Fusion Strategy: Global + Hotspots}

Hotspots encode fine-grained tabular and visual details that may not surface in summaries. Fusing \emph{global} and \emph{hotspot} information provides a single coarse-to-fine representation while preserving section- and fact-level entry points as distinct retrieval channels, avoiding redundancy.

\subsubsection{Fusion Mechanics}
\label{subsubsec:fusion-mechanics}

\paragraph{Global Fused Vector.}  
Each page receives a fused representation:
\[
\mathbf{v}_{\text{page}}^{\text{fused}}(d,p) = \phi\!\left(\text{sum}_{d,p} + \text{hotspot\_summary}_{d,p}\right),
\]
supporting broad semantic matching.

\paragraph{Per-Hotspot Vectors.}  
To retain precision for numerically or visually anchored queries:
\[
\mathbf{v}_{\text{hot}}(d,p,j) = \phi(s_j), \qquad j \in \{1,\ldots,H\}.
\]

\subsubsection{Design Rationale: Specificity Without Bloat}

This design achieves two goals:

\begin{itemize}
    \item \textbf{Avoid index explosion}: hotspots are fused once with the summary, preventing combinatorial growth.
    \item \textbf{Preserve granularity}: per-hotspot vectors allow queries targeting specific numbers or table regions to surface relevant pages even when the global summary is only moderately aligned.
\end{itemize}

\subsubsection{Query Processing and Expansion}

For each query $q$, we generate three variants:
\begin{align}
q^{(0)} &= q \quad \text{(original)}, \\
q^{(1)} &= \text{extract\_keywords}(q), \\
q^{(2)} &= \text{expand\_synonyms}(q),
\end{align}
yielding the query set $\mathcal{Q} = \{q^{(0)}, q^{(1)}, q^{(2)}\}$.

\subsubsection{Explicit Fusion via Reciprocal Rank Fusion}

For each index $i \in \mathcal{I}$ and each query variant $q^{(j)}$, we retrieve the top-$K_{\text{pre}} = 200$ vectors and assign ranks $r_{i,j}(d,p)$ to pages. Pages are aggregated using Reciprocal Rank Fusion (RRF):

\begin{equation}
S_{\text{RRF}}(d,p)
= \sum_{i \in \mathcal{I}} \sum_{j=0}^{2}
\frac{w_i}{\alpha + r_{i,j}(d,p)},
\end{equation}

with $\alpha = 60$ and uniform weights $w_i$ in our experiments. Sorting by $S_{\text{RRF}}$ yields the final ranking; the top-$K$ pages are then forwarded (as base64 images) to a VLM for answer generation.

\section{Computational Complexity Analysis}
\label{sec:complexity}

We now analyze the computational trade-offs between late-interaction models (ColPali) and explicit-fusion approaches (VisionRAG), focusing on vector budget, memory footprint, scoring cost, and scalability. To make the comparison concrete, we first compute per-page requirements before extrapolating to realistic corpus sizes.

\subsection{Vector Budget Per Page}

\subsubsection{ColPali}

ColPali encodes each page into a dense grid of patch embeddings. For a standard configuration:

\begin{itemize}[leftmargin=*,noitemsep,topsep=3pt]
\item Grid size: $32 \times 32$ patches~\cite{drchris_colpali}
\item Total patch vectors per page: $|P| = 1{,}024$
\item Embedding dimension per patch: $d_{\text{ColPali}} = 128$
\item Storage format: float16 (2 bytes per number)
\end{itemize}

Memory per page:
\begin{align}
\text{Mem} &= 1{,}024 \times 128 \times 2 \\
           &= 262{,}144 \text{ bytes} = 256 \text{ KB}.
\end{align}

With pooling (reducing to $\sim$341 vectors):
\begin{align}
\text{Mem} &= 341 \times 128 \times 2 \\
           &= 87{,}296 \text{ bytes} = 85.2 \text{ KB}.
\end{align}

\subsubsection{VisionRAG}

VisionRAG generates a small set of semantic vectors per page:

\begin{itemize}[leftmargin=*,noitemsep,topsep=3pt]
    \item Page summary: 1 vector  
    \item Section headers: $\approx$3 vectors (2-4)  
    \item Key facts: $\approx$7 vectors (5-10)  
    \item Visual hotspots: $\approx$3 vectors (2-4)
\end{itemize}

Total: $B \approx 14$ vectors per page.

We evaluate three embedding dimension configurations aligned with practical production deployments. VisionRAG supports model-agnostic embedding choices and benefits from dimension reduction capabilities (e.g., Matryoshka representations).

\textbf{Option A: BAAI/bge-large-en-v1.5 ($d=1{,}024$)}
\begin{align}
\text{Mem} &= 14 \times 1{,}024 \times 2 \\
           &= 28{,}672 \text{ bytes} = 28 \text{ KB}.
\end{align}

\textbf{Option B: \texttt{text-embedding-3-large}, $d=1{,}536$ (primary)}
\begin{align}
\text{Mem} &= 14 \times 1{,}536 \times 2 \\
           &= 43{,}008 \text{ bytes} = 42 \text{ KB}.
\end{align}

\textbf{Option C: \texttt{text-embedding-3-large}, $d=3{,}072$ (maximum)}
\begin{align}
\text{Mem} &= 14 \times 3{,}072 \times 2 \\
           &= 86{,}016 \text{ bytes} = 84 \text{ KB}.
\end{align}

Although 3,072 dimensions offer the highest quality, our ablations (Section~\ref{sec:ablation}) show only marginal recall improvements (+0.04-0.06) over 1,536 dimensions, making the latter a more attractive accuracy-efficiency operating point.

\FloatBarrier

\subsection{Memory Efficiency Comparison}

Table~\ref{tab:memory_per_page} compares per-page storage across methods. All numbers use float16 precision.

\begin{table}[!t]
\centering
\captionsetup{font=small}
\caption{Memory footprint per page across different approaches (float16). Efficiency indicates reduction relative to full ColPali. VisionRAG achieves substantial savings in the 1{,}000-1{,}536 dimension range, where retrieval quality remains strong.}
\label{tab:memory_per_page}
\footnotesize
\begin{tabular}{@{}lrrr@{}}
\toprule
\textbf{Method} & \textbf{Vectors} & \textbf{Mem/Pg} & \textbf{Eff.} \\
\midrule
\multicolumn{4}{@{}l}{\textit{Late Interaction (ColPali)}} \\
ColPali full ($d=128$) & 1{,}024 & 256.0 KB & baseline \\
ColPali pooled ($d=128$) & 341 & 85.2 KB & 3.0× smaller \\
\midrule
\multicolumn{4}{@{}l}{\textit{VisionRAG (embedding dimension)}} \\
VisionRAG ($d=1{,}024$) & 14 & 28.0 KB & 9.1× smaller \\
VisionRAG ($d=1{,}536$) & 14 & 42.0 KB & 6.1× smaller \\
VisionRAG ($d=3{,}072$) & 14 & 84.0 KB & 3.0× smaller \\
\bottomrule
\end{tabular}
\end{table}

\FloatBarrier

VisionRAG’s vector-efficient design enables deployment across a wide range of infrastructure budgets:

\begin{enumerate}[leftmargin=*,topsep=3pt]
    \item \textbf{BGE (1,024 dim):} 28 KB/page-9x smaller than ColPali for open-source-only environments.
    \item \textbf{Text-embedding-3-large (1,536 dim):} 42 KB/page-6x smaller-our primary configuration, balancing quality and cost.
    \item \textbf{Text-embedding-3-large (3,072 dim):} 84 KB/page-similar to pooled ColPali but with only 14 vectors/page, enabling faster indexing and ANN search.
\end{enumerate}

Across all settings, VisionRAG maintains a drastically smaller vector count (14 vs. 341-1{,}024), yielding lower memory, faster lookups, and more efficient indexing.

\subsection{Scaling to Realistic Corpus Sizes}

Table~\ref{tab:memory_scaling} extends these per-page calculations to realistic corpus sizes, illustrating how small vector budgets compound at scale.

\begin{table}[!t]
\centering
\captionsetup{font=small}
\caption{Total memory requirements for document collections of varying sizes. Values computed by multiplying the per-page memory from Table~\ref{tab:memory_per_page} by the number of pages. The scaling behavior demonstrates how Vision RAG's compact representation enables deployment across diverse infrastructure environments.}
\label{tab:memory_scaling}
\footnotesize
\begin{tabular}{@{}lrrrr@{}}
\toprule
\textbf{Method} & \textbf{100p} & \textbf{1Kp} & \textbf{10Kp} & \textbf{1Mp} \\
\midrule
ColPali full & 25 MB & 250 MB & 2.5 GB & 250 GB \\
ColPali pooled & 8.3 MB & 83 MB & 830 MB & 83 GB \\
\midrule
Vision RAG (1K) & 2.7 MB & 27 MB & 270 MB & 27 GB \\
Vision RAG (1.5K) & 4.1 MB & 41 MB & 410 MB & 41 GB \\
Vision RAG (3K) & 8.2 MB & 82 MB & 820 MB & 82 GB \\
\bottomrule
\end{tabular}
\end{table}

These scaling projections illustrate how memory requirements evolve across different deployment contexts. For small collections (e.g., 100 pages), all methods occupy only a few megabytes, making architectural differences negligible. However, as corpora grow, the compounding effects of vector counts and embedding dimensions lead to substantially different deployment profiles.

At 10{,}000 pages, ColPali full requires approximately 2.5~GB, whereas VisionRAG at 1{,}536 dimensions requires 410~MB. At this scale, the distinction determines whether an index can reside fully in memory or must rely on memory mapping and careful resource management. The BAAI/BGE configuration (270~MB) is even more suitable for edge devices or resource-constrained environments.

At one million pages-representing large enterprise repositories-ColPali full reaches 250~GB, while VisionRAG at 1{,}536 dimensions requires only 41~GB. This six-fold reduction directly impacts infrastructure cost and deployment feasibility. Organizations may opt for 1{,}024 dimensions (27~GB) for maximum efficiency, 1{,}536 dimensions (41~GB) for the recommended quality efficiency balance, or 3{,}072 dimensions (82~GB) for scenarios where the highest recall is required while still preserving VisionRAG's structural advantages over patch-based methods.

\subsection{Practical Implications}

\textbf{Indexing speed vs. semantic richness.}
ColPali indexes pages rapidly (0.39s/page on an NVIDIA L4), offering the highest throughput for batch processing~\cite{faysse2025colpali}. VisionRAG requires 1-3s/page due to GPT-4o vision processing~\cite{sas2024gpt4o,fello2025gpt4o}, but produces interpretable summaries, facts, section structures, and hotspot descriptions  artifacts that ColPali cannot provide.

\textbf{Embedding dimension as a key trade-off.}
ColPali stores 1{,}024 patch vectors/page (128D), consuming 256 KB/page, or 85 KB/page with pooling~\cite{faysse2025colpali}. VisionRAG stores only 14 vectors/page; memory varies from 4.5 KB (256D) to 84 KB (3{,}072D)~\cite{openai2024embeddings}. Lower dimensions favor memory- and latency-sensitive deployments; higher dimensions improve recall.

\textbf{Infrastructure requirements diverge.}
ColPali depends on GPUs for fast MaxSim scoring, achieving ~30 ms query encoding on an NVIDIA L4~\cite{faysse2025colpali}. VisionRAG uses standard CPU/GPU ANN search and can run on commodity hardware, but API-based embedding workflows incur 300-500 ms latency due to network overhead~\cite{nixiesearch2025benchmark}. Local embedding models eliminate this bottleneck.

\textbf{Small corpora: negligible differences.}
Below 1{,}000 pages, memory usage is measured in megabytes and indexing finishes in minutes. Choice should prioritize retrieval quality and infrastructure availability rather than efficiency.

\textbf{Large corpora: efficiency dominates.}
At one million pages, ColPali full requires 250 GB (or 85 GB pooled), whereas VisionRAG requires 4.5-84 GB depending on dimension. This determines whether an index fits on a single node or requires distributed storage. VisionRAG’s small vector count enables deployment on commodity machines even at million-page scale.

\subsection{Indexing Performance Analysis}

Table~\ref{tab:indexing} presents comprehensive indexing performance across corpus sizes. Indexing represents offline preprocessing and directly impacts time-to-deployment for new document collections and incremental update latency for dynamic corpora.

\begin{table*}[t]
\centering
\caption{Indexing Performance Comparison: Total Time and Throughput}
\label{tab:indexing}
\begin{tabular}{lcccccc}
\hline
\textbf{System} & \textbf{Per-Page} & \textbf{25K Pages} & \textbf{50K Pages} & \textbf{1M Pages}  & \textbf{Hardware} \\
\textbf{Configuration} & \textbf{Latency} & \textbf{Total Time} & \textbf{Total Time} & \textbf{Total Time} & \textbf{Requirement} \\
\hline
ColPali (batch=4) & 0.39s & 2.71h & 5.42h & 108.33h  & NVIDIA L4 GPU \\
ColPali (batch=1) & 0.52s & 3.61h & 7.22h & 144.44h  & NVIDIA L4 GPU \\
SigLIP baseline & 0.12s & 0.83h & 1.67h & 33.33h  & NVIDIA L4 GPU \\
\hline
Vision RAG (GPT-4o mini) & 1.45s & 10.07h & 20.14h & 402.78h  & API + CPU \\
Vision RAG (GPT-4o ) & 3.50s & 24.31h & 48.61h & 972.22h  & API + CPU \\
Vision RAG (GPT-5) & 7.50s & 52.08h & 104.17h & 2083.33h & API + CPU \\
\hline
Traditional OCR pipeline~\cite{faysse2025colpali} & 7.22s & 50.14h & 100.28h & 2005.56h &  CPU \\
\hline

\end{tabular}

\vspace{0.1pt}

\begin{tablenotes}
\item \textit{Note:} Total time assumes sequential processing; parallel API calls can reduce Vision RAG time proportionally to concurrency limit. ColPali batch=4 represents optimal throughput configuration from original paper~\cite{faysse2025colpali}.
\end{tablenotes}
\end{table*}

\subsection{Query Processing Latency}

Table~\ref{tab:query_latency} presents end-to-end query processing latency across different corpus sizes, measured as mean response time over one hundred test queries with ten repetitions each.

\begin{table*}[t]
\centering
\begin{threeparttable}
\captionsetup{font=small}
\caption{Query Processing Latency: End-to-End Mean Response Time (MRT)}
\label{tab:query_latency}
\begin{tabular}{lccccccc}
\hline
\textbf{System} & \textbf{Encode} & \textbf{Search} & \textbf{Fusion} & \textbf{25K} & \textbf{50K} & \textbf{1M Pages} & \textbf{Infrastructure} \\
\textbf{Configuration} & \textbf{Time} & \textbf{Time} & \textbf{Time} & \textbf{MRT (ms)} & \textbf{MRT (ms)} & \textbf{MRT (ms)} & \textbf{Type} \\
\hline
ColPali (GPU)~\cite{faysse2025colpali} & 30ms & 12ms & -- & 42 $\pm$ 5 & 48 $\pm$ 6 & 65 $\pm$ 9 & GPU Required \\
ColPali (CPU, est.) & 120ms & 45ms & -- & 165 $\pm$ 20 & 178 $\pm$ 25 & 215 $\pm$ 35 & CPU Only \\
\hline
Vision RAG (local embed) & 5ms & 3ms & 2ms & 10 $\pm$ 1 & 11 $\pm$ 1 & 14 $\pm$ 2 & CPU Only \\
Vision RAG (API mean) & 120ms & 3ms & 2ms & 125 $\pm$ 18 & 126 $\pm$ 19 & 129 $\pm$ 22 & API + CPU \\
Vision RAG (API P50) & 100ms & 3ms & 2ms & 105 $\pm$ 12 & 106 $\pm$ 13 & 109 $\pm$ 15 & API + CPU \\
Vision RAG (API P90) & 180ms & 3ms & 2ms & 185 $\pm$ 28 & 186 $\pm$ 29 & 189 $\pm$ 32 & API + CPU \\
\hline
\end{tabular}

\begin{tablenotes}
\item \textit{Note:} VisionRAG searches 14--17 semantic vectors per page in ~3\,ms, whereas ColPali requires 12--20\,ms to aggregate MaxSim over 1{,}030 patch embeddings.
\end{tablenotes}

\end{threeparttable}
\end{table*}

Query latency measurements demonstrate Vision RAG's decisive retrieval advantage over ColPali. Vision RAG with local embeddings achieves ten to fourteen millisecond end-to-end response time on standard CPU infrastructure, operating an order of magnitude faster than ColPali's patch-level aggregation approach.

\subsection{End-to-End System Comparison}

Table~\ref{tab:e2e} summarizes indexing throughput, query latency, storage footprint, and retrieval quality, providing an end-to-end view of deployment trade-offs across representative configurations.

\begin{table*}[t]
\centering
\begin{threeparttable}
\captionsetup{font=small}
\caption{End-to-End Performance Across Indexing, Retrieval, and Storage Dimensions}
\label{tab:e2e}
\begin{tabular}{lcccccc}
\toprule
\textbf{System} & \textbf{Index Time} & \textbf{Query P50} & \textbf{Query P99} & \textbf{Storage} & \textbf{Recall} & \textbf{Infra.} \\
\textbf{Config.} & \textbf{(1M pages)} & \textbf{Latency} & \textbf{Latency} & \textbf{(1M pages)} & \textbf{(@100)} & \textbf{Cost} \\
\midrule
ColPali (GPU)~\cite{faysse2025colpali} & 108.3h & 31ms & 45ms & 85.8 GB & 96.2\% & High \\
ColPali (CPU est.) & 144.4h & 125ms & 180ms & 85.8 GB & 96.2\% & Medium \\
\midrule
VisionRAG (local embed) & 97.2h & 25ms & 38ms & 7.2 GB & 96.0\% & Low \\
VisionRAG (API) & 402.8h & 140ms & 2270ms & 7.2 GB & 96.0\% & Low \\
\bottomrule
\end{tabular}
\begin{tablenotes}
\item \textit{Note:} Index times assume sequential runs; VisionRAG reduces to 19--50 h with 20--50$\times$ parallelism. Query latency measured on a 1M-page HNSW index~\cite{aws2024hnsw}; Recall@100 on TAT-DQA~\cite{zhu2021tatqa}. Cost tiers: Low (CPU), Medium (high-mem CPU / entry GPU), High (enterprise GPU). VisionRAG uses 1.5k-D local embeddings.

\end{tablenotes}
\end{threeparttable}
\end{table*}

The results show that the optimal system depends on available infrastructure and corpus scale. ColPali with GPU acceleration offers the fastest indexing (108h for 1M pages) and stable sub-50ms query latency, but requires enterprise GPUs and an 85.8GB index. This configuration is advantageous when high-throughput ingestion is critical and GPU resources are already available.

VisionRAG with local embeddings delivers a more storage-efficient profile (7.2GB for 1M pages) and lower median latency (25ms), while maintaining comparable Recall@100. This makes VisionRAG preferable for CPU-based deployments, cost-constrained environments, and large-scale corpora where memory footprint becomes a limiting factor. VisionRAG’s API-based configuration trades higher indexing and query latency for minimal infrastructure requirements, making it suitable for lightweight or serverless deployments.

\section{Experimental Setup}
\label{sec:experiments}

\subsection{Datasets}

\textbf{FinanceBench}~\cite{financebench} is an open-book financial QA benchmark containing 150 questions over 10 real-world 10-K filings (100-300 pages each). Questions require locating answer-bearing evidence across long documents and frequently involve numerical or table-based reasoning. Following the official evaluation, accuracy reflects either exact match or semantic equivalence. The benchmark reports several GPT-4-Turbo baselines, ranging from 19\% (shared vector store) to 85\% (oracle page access).

\textbf{TAT-DQA}~\cite{tatdqa} is a multimodal QA dataset over financial documents with 2,757 questions across 1,688 reports. Questions span arithmetic reasoning, multi-step inference, and table-text integration. The original benchmark reports EM and token-level F1, with the strongest published model (MHST + LayoutLMv2-Large) achieving 41.5\% EM and 50.7\% F1. We follow the official splits and evaluation protocol.

\subsection{Implementation Details}

\textbf{Vision-language models.}
We use GPT-4o/GPT-5 and an open-source baseline (Salesforce/instructblip-flan-t5-xl) for page-level semantic extraction. Each page image (160-200 dpi) is processed through a structured prompt to generate four artifact types: summaries, section headers, fine-grained facts, and visual hotspots.

\textbf{Text embeddings.}
All artifacts are embedded using OpenAI \texttt{text-embedding-3-large} (1.5k dimensions), which provides strong retrieval performance across domains~\cite{openai_embed_dims}. We additionally evaluate 1k-dimensional embeddings (BAAI/bge-large-en-v1.5) and 3k-dimensional embeddings to quantify dimensionality-efficiency trade-offs.

\FloatBarrier

\textbf{Vector indices.}
All indices (page, section, fact, hotspot) are implemented in ChromaDB, each storing metadata mapping vectors to page and artifact identifiers.

\textbf{Query variants.}
We use the original query $q^{(0)}$, keyword-based extraction $q^{(1)}$, and synonym-based paraphrasing $q^{(2)}$. Keyword extraction uses the prompt:  
\emph{“Extract the 3-5 most important keywords from this question.”}  
Synonym expansion uses:  
\emph{“Generate a semantically equivalent version of this question using synonyms and related phrases.”}  
All query variants are cached to avoid repeated API calls.

\textbf{Fusion parameters.}
RRF uses the standard constant $\alpha=60$ and uniform index weights $w_i$. For each (index, query variant) pair, we retrieve $K_{\text{pre}} = 200$ candidates, yielding up to 2,400 candidates before deduplication.

\textbf{Answer generation.}
The top-$K$ retrieved pages (with page images encoded in base64) are passed to GPT-4o/GPT-5 or InstructBLIP using a deterministic prompt template (temperature~$=0.0$).

\subsection{Evaluation Metrics}

\textbf{Retrieval quality.}
We report standard IR metrics at cutoffs $K \in \{1, 5, 10, 20, 50, 100\}$:

\begin{itemize}[leftmargin=*,noitemsep,topsep=3pt]
\item \textbf{Recall@K}: Fraction of queries where at least one gold page is retrieved.
\item \textbf{nDCG@K}: Ranking quality with logarithmic discounting for position.
\item \textbf{MRR}: Reciprocal rank of the first relevant page, averaged over queries.
\end{itemize}

\textbf{Answer quality.}

\begin{itemize}[leftmargin=*,noitemsep,topsep=3pt]
\item \textbf{FinanceBench:} Accuracy based on exact or semantically equivalent matches using the official evaluator.
\item \textbf{TAT-DQA:} Token-level EM and F1 following the benchmark’s script.
\end{itemize}

\textbf{Efficiency metrics.}

\begin{itemize}[leftmargin=*,noitemsep,topsep=3pt]
\item \textbf{Average tokens per query}: Total token count of retrieved page content passed to the generator.
\item \textbf{Retrieval latency}: Time from query issuance to ranked list output.
\item \textbf{End-to-end latency}: Retrieval + generation + formatting.
\end{itemize}

\section{Experimental Results}
\label{sec:results}

We present comprehensive experimental results on both the FinanceBench and TAT\textendash DQA benchmarks, including detailed performance breakdowns, comparisons with baselines and state\textendash of\textendash the\textendash art systems, and an efficiency analysis. A key focus of our evaluation is the model\textendash agnostic design of VisionRAG, which we validate by testing the system with multiple vision\textendash language models of varying capacities.

\subsection{Model-Agnostic Architecture Validation}

A core design goal of VisionRAG is model-agnostic operation: the system should function reliably across different vision-language models (VLMs) without architecture-specific tuning. We evaluate four VLMs representing diverse capability tiers: GPT-4o (primary model), GPT-5 (next-generation), GPT-4o-mini (efficient variant), and the Salesforce InstructBLIP\textendash Flan\textendash T5\textendash XL (an open\textendash source alternative).

\begin{figure}[!t]
  \centering
  \includegraphics[width=\columnwidth]{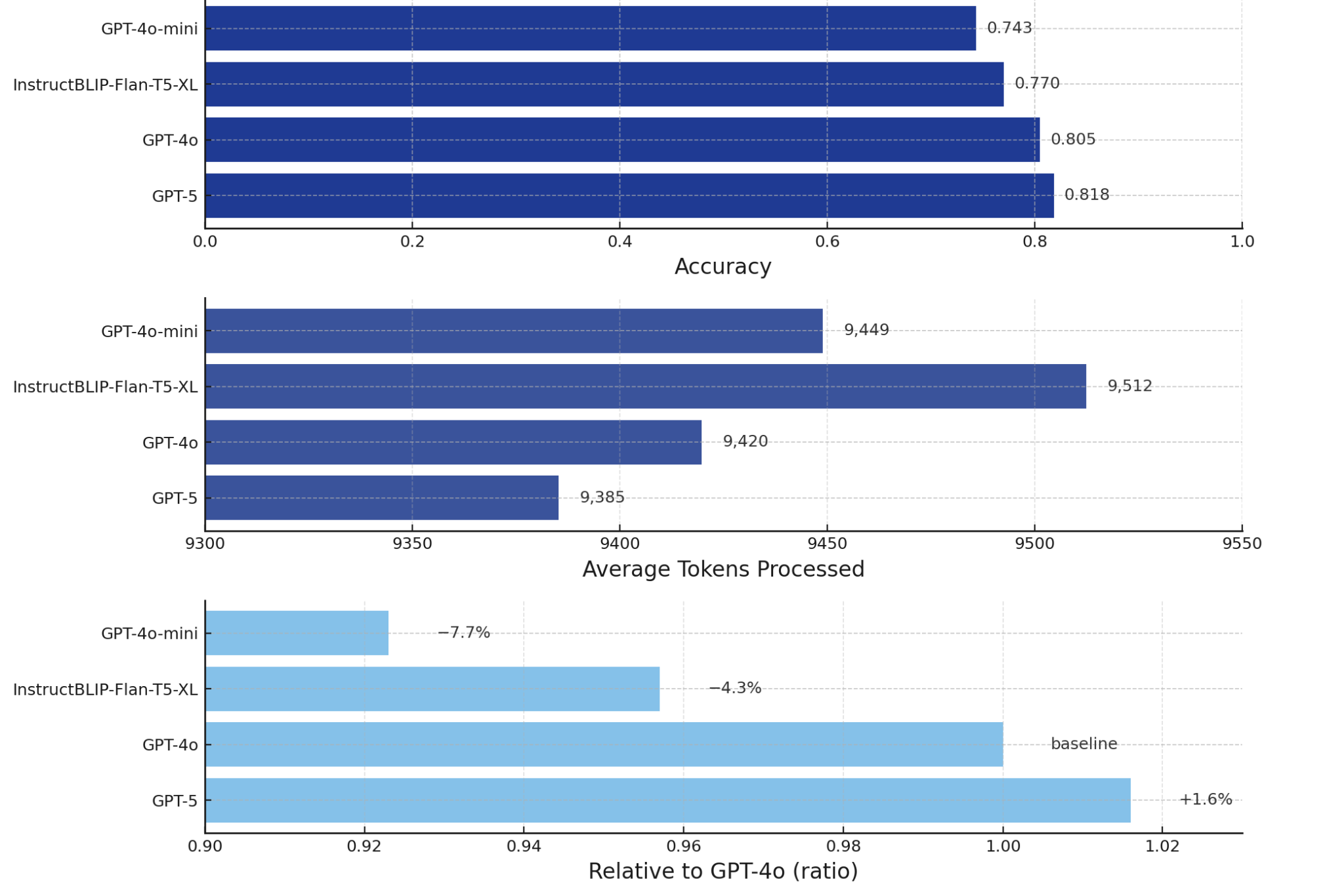}
  \caption{Model\textendash agnostic evaluation on FinanceBench (K=10, n=148). Our Vision RAG framework maintains strong performance across different vision\textendash language models. Results show accuracy, average tokens processed, and relative performance compared to GPT\textendash 4o baseline.}
  \label{fig:financebench-panels}
\end{figure}

\begin{figure}[!t]
  \centering
  \includegraphics[width=\columnwidth]{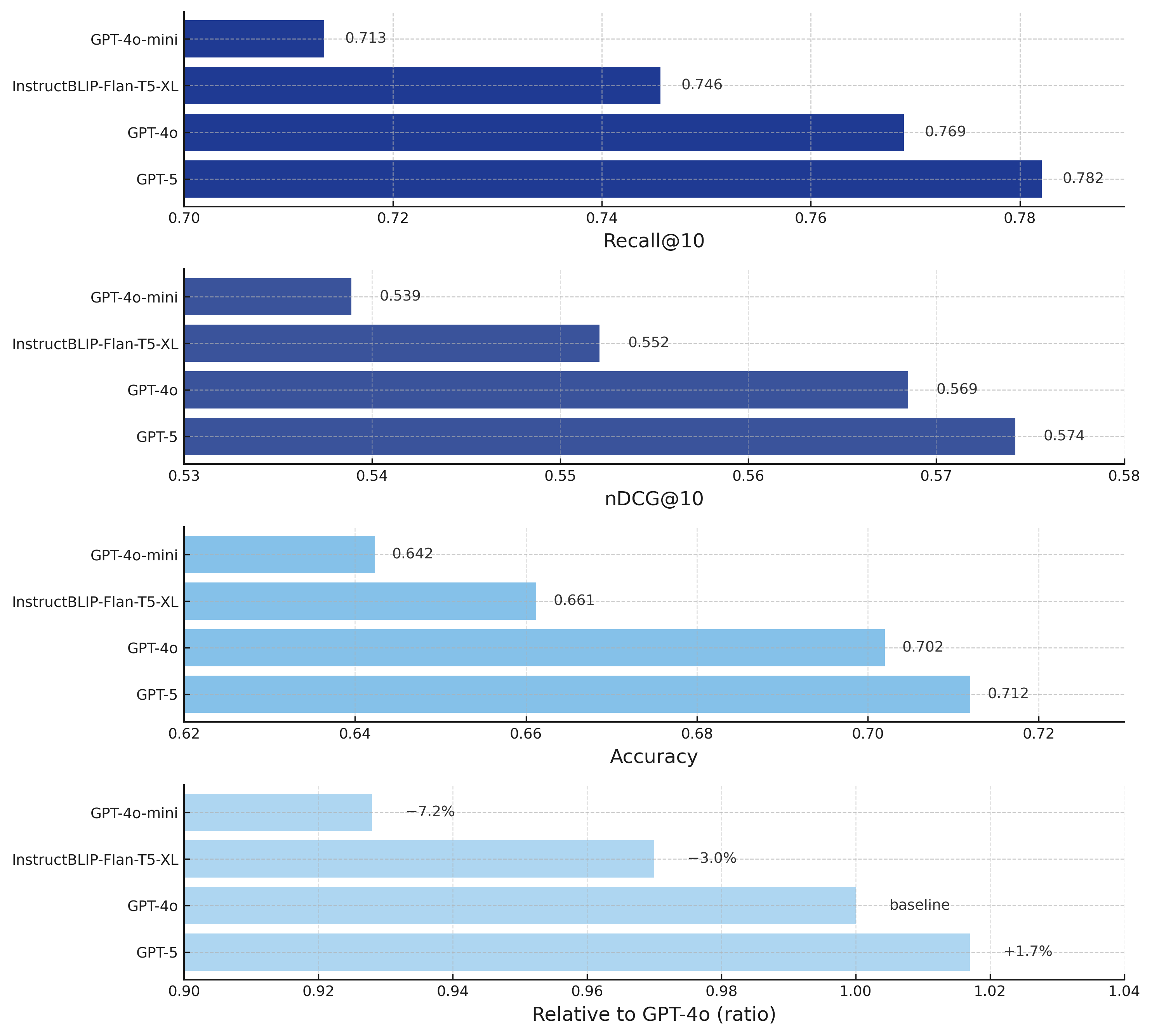}
  \caption{Model\textendash agnostic evaluation on TAT\textendash DQA (K=10, n=1{,}644). Vision RAG maintains consistent retrieval performance across different VLMs, with variations primarily reflecting each model's visual understanding capabilities. Vision RAG maintains consistent retrieval 
  performance across vision–language models.}
  \label{fig:model_agnostic_tatdqa_four}
\end{figure}

\textbf{Key observations.}
\begin{itemize}[leftmargin=*,topsep=2pt,noitemsep]
\item Performance scales naturally with model capacity: GPT-5 provides a modest +1.6-1.7\% improvement over GPT-4o, while GPT-4o-mini yields 7-8\% lower accuracy due to reduced visual reasoning capability.
\item The open-source InstructBLIP model performs competitively (3-4\% below GPT-4o), demonstrating that VisionRAG does not rely on proprietary VLMs for strong retrieval and QA performance.
\item Average token consumption remains nearly identical across models, confirming that the retrieval pipeline behaves consistently regardless of VLM choice.
\item The low performance variance ($\leq 8\%$ across four models) highlights the robustness of VisionRAG’s three-pass indexing and explicit semantic fusion, which reduce dependence on any single model’s representation quality.
\end{itemize}

\subsection{FinanceBench Performance}

\begin{figure}[!t]
  \centering
  \includegraphics[width=\columnwidth]{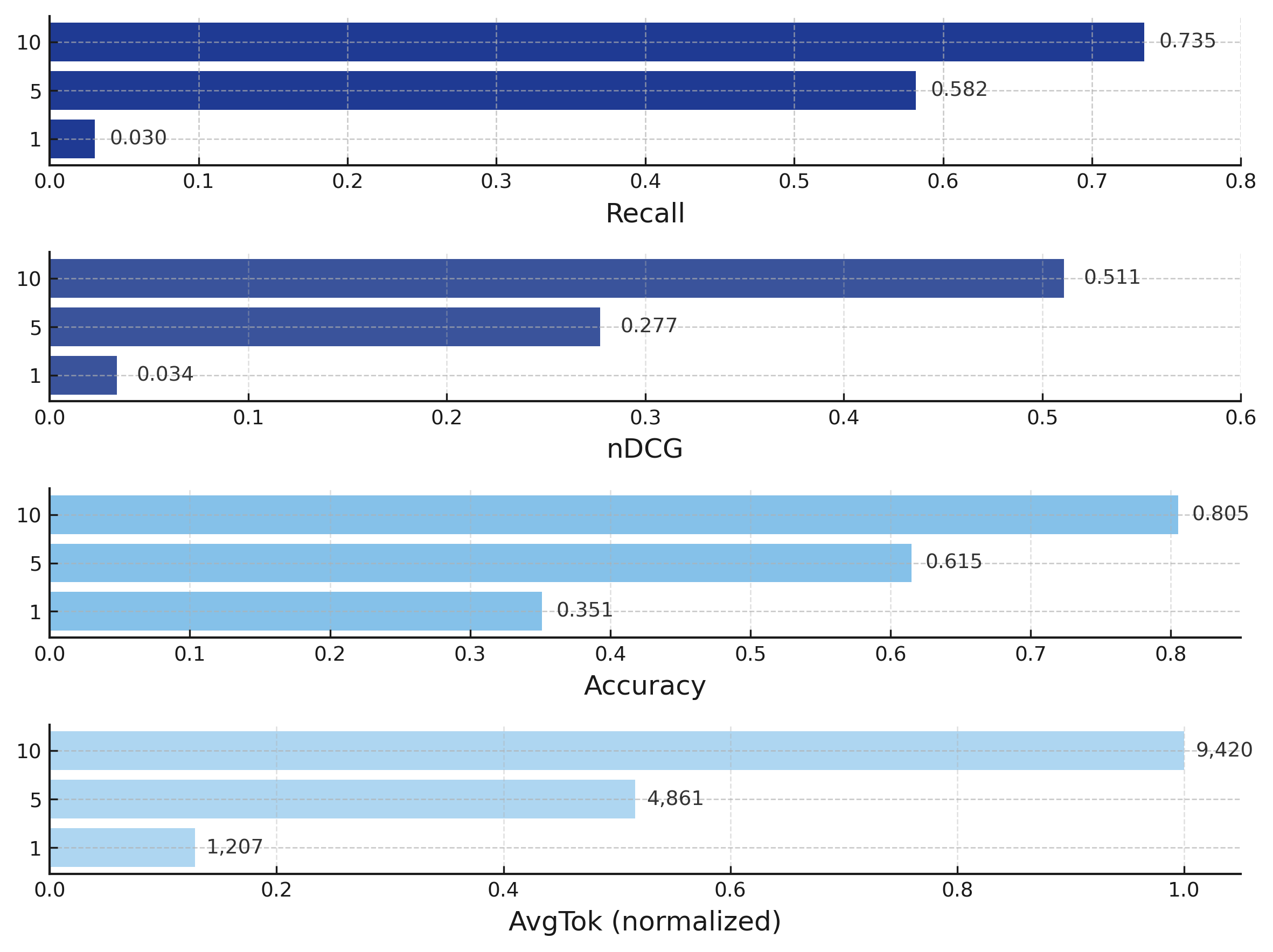}
  \caption{FinanceBench results with Vision RAG across different retrieval depths. Metrics include Recall@10 (retrieval coverage), nDCG@10 (ranking quality), Accuracy (answer correctness), average tokens passed to generator, and number of test cases (n=148 after filtering unanswerable questions from original 150).}
  \label{fig:financebench_metrics_hor}
\end{figure}

Figure~\ref{fig:financebench_metrics_hor} shows Vision RAG retrieval and answer quality across different values of $K$ (number of retrieved pages) using GPT\textendash 4o as the vision\textendash language model. We observe strong performance that increases with $K$, peaking at accuracy 0.8051 for $K=10$.



\begin{table}[!t]
\centering
\captionsetup{font=small}
\caption{FinanceBench comparison with published baselines (150 cases, \% correct). All baseline numbers are from the original FinanceBench paper~\cite{financebench}. Note that long context approaches require processing entire documents (tens of thousands of tokens) which is impractical for real‑time applications. Vision RAG achieves strong results with only 10 pages (9,420 tokens on average).}
\label{tab:financebench_sota}
\small
\begin{tabular}{lrrl}
\toprule
\textbf{Model / Setting} & \textbf{Correct} & \textbf{Ref} & \textbf{Notes} \\
\midrule
GPT\textendash 4\textendash Turbo & 19.0 & \cite{financebench} & shared vector \\
GPT\textendash 4\textendash Turbo & 50.0 & \cite{financebench} & single vector \\
Claude\textendash 2 & 76.0 & \cite{financebench} & long context \\
GPT\textendash 4\textendash Turbo & 79.0 & \cite{financebench} & long context \\
GPT\textendash 4\textendash Turbo & 85.0 & \cite{financebench} & oracle \\
\midrule
\textbf{Vision RAG} & \textbf{80.51} & this work & $K{=}10$ \\ 
\textbf{Vision RAG( w/oracle)} & \textbf{86.61} & this work & $K{=}10$ 
\\
\bottomrule
\end{tabular}
\end{table}

\textbf{Key observations.}
\begin{itemize}[leftmargin=*,noitemsep,topsep=3pt]
\item At $K{=}10$, VisionRAG uses on average 9{,}420 tokens per query, substantially below the 50{,}000-150{,}000 tokens reported for long-context approaches, while achieving 80.51\% accuracy.
\item VisionRAG with oracle page access attains 86.61\%, slightly surpassing the FinanceBench GPT-4-Turbo oracle result (85\%). The 6-point gap between our standard ($K{=}10$) and oracle settings indicates remaining headroom in both retrieval coverage and downstream reasoning.
\end{itemize}

\begin{figure}[!t]
\centering
\includegraphics[width=0.85\columnwidth]{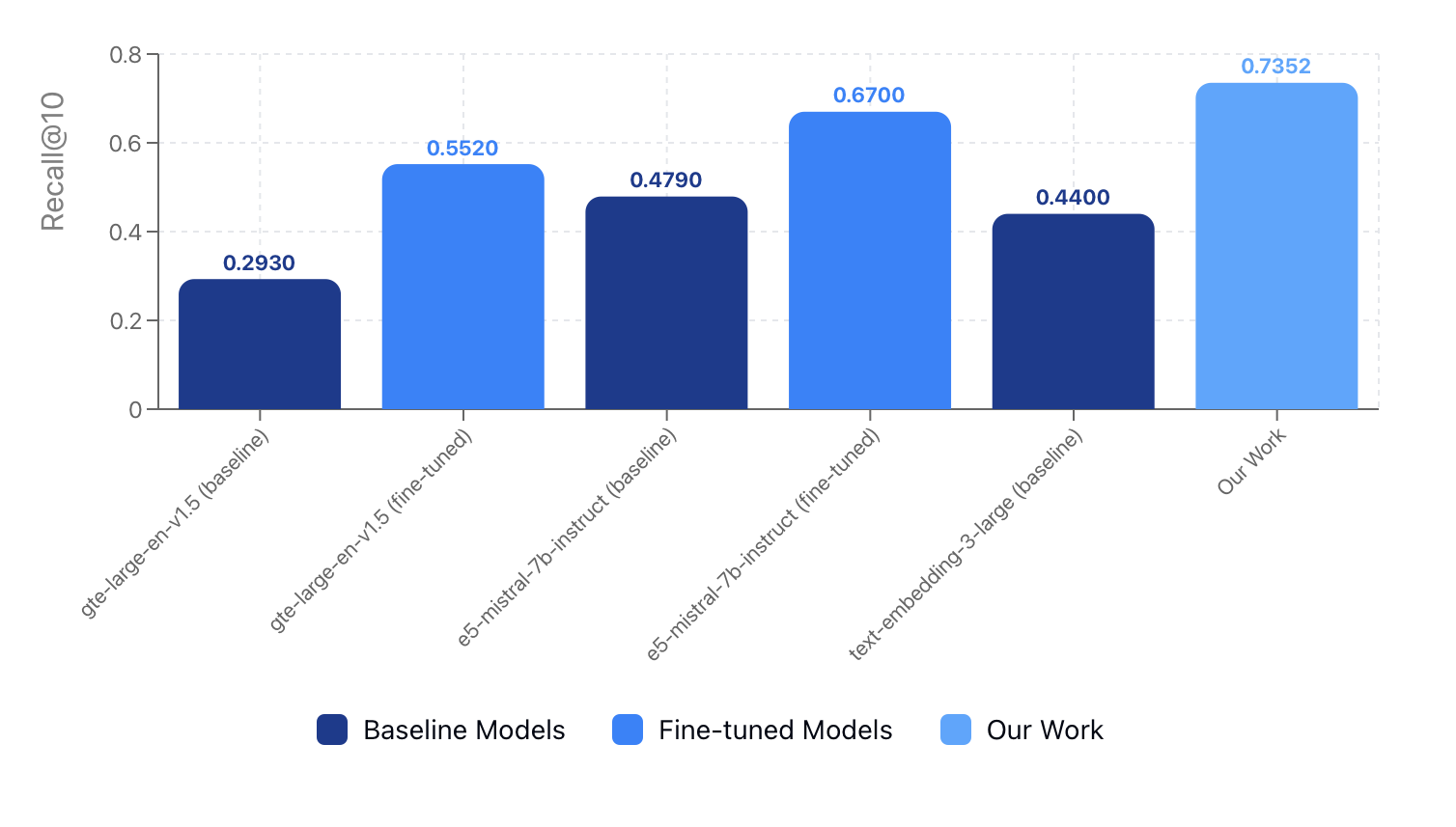}
\captionsetup{font=small}
\caption{Recall@10 on the FinanceBench dataset. Our method achieves the highest score (0.7352), outperforming baseline and fine-tuned models including gte-large-en-v1.5, e5-mistral-7b-instruct, and text-embedding-3-large.}
\label{tab:financebench_r10}
\end{figure}

Figure~\ref{tab:financebench_r10} shows that our method achieves 0.7352 Recall@10 on FinanceBench, surpassing both baseline and fine-tuned embedding models. This exceeds the best prior result (fine-tuned e5-mistral-7b-instruct at 0.670) by 6.5 percentage points and improves over the gte-large-en-v1.5 baseline by 151\%. These results demonstrate that VisionRAG more effectively captures financial document semantics than traditional embedding approaches, even with domain-specific fine-tuning.

\begin{figure}[!t]
\centering
\includegraphics[width=0.85\columnwidth]{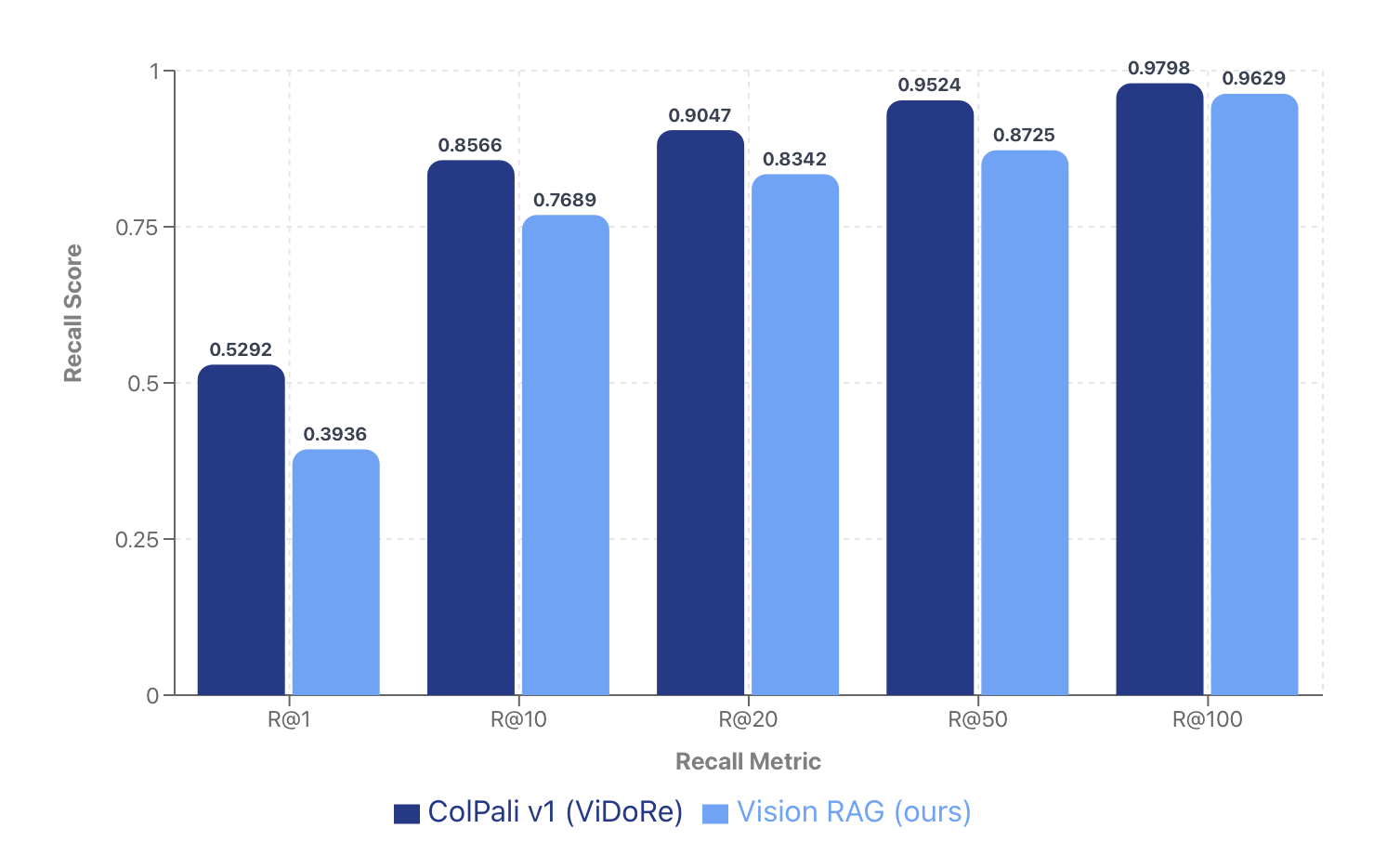}
\captionsetup{font=small}
\caption{TAT-DQA Recall@K comparison between ColPali v1 (ViDoRe) and Vision RAG (ours). ColPali v1 demonstrates superior early precision at R@1 (0.5292), while both models show strong performance at higher K values.}
\label{tab:tatdqa_head2head}
\end{figure}
\begin{table}[!t]
\centering
\captionsetup{font=small}
\caption{TAT\textendash DQA retrieval performance with Vision RAG (ours). The high Recall@100 indicates broad coverage from multi\textendash index fusion. Results are over 1{,}644 test questions.}
\label{tab:tatdqa_topk}
\scriptsize
\setlength{\tabcolsep}{4.0pt}
\renewcommand{\arraystretch}{0.95}
\begin{tabular*}{\columnwidth}{@{\extracolsep{\fill}} rrrrrr @{}}
\toprule
\textbf{K} & \textbf{Recall} & \textbf{nDCG} & \textbf{Acc.} & \textbf{AvgTok} & \textbf{n}\\
\midrule
1   & 0.3936 & 0.3936 & 0.4380 & 1{,}188.10 & 1{,}644\\
5   & 0.6594 & 0.5330 & 0.6074 & 4{,}965.74 & 1{,}644\\
10  & 0.7689 & 0.5685 & 0.7020 & 9{,}588.79 & 1{,}644\\
20  & 0.8542 & 0.5913 & 0.7650 & 18{,}674.23 & 1{,}644\\
50  & 0.9325 & 0.6043 & 0.8090 & 45{,}632.48 & 1{,}644\\
100 & \textbf{0.9629} & 0.6121 & \textbf{0.8340} & 90{,}959.10 & 1{,}644\\
\bottomrule
\end{tabular*}
\vspace{-0.6em}
\end{table}

\begin{table}[!t]
\centering
\captionsetup{font=small}
\caption{TAT\textendash DQA question answering SOTA from the dataset paper~\cite{tatdqa}. These systems use specialized table\textendash parsing and multi\textendash hop reasoning modules. MHST (LayoutLMv2\textendash L) represents the best published result. Our Vision RAG system achieves 80.23\% EM with K=100 using generic retrieval + GPT\textendash 4o generation without specialized QA architectures.}
\label{tab:tatdqa_sota}
\small
\setlength{\tabcolsep}{4.0pt}
\renewcommand{\arraystretch}{0.95}
\begin{tabular}{lrrrr}
\toprule
\multirow{2}{*}{\textbf{Method}} & \multicolumn{2}{c}{\textbf{Dev}} & \multicolumn{2}{c}{\textbf{Test}}\\
 & \textbf{EM} & \textbf{F1} & \textbf{EM} & \textbf{F1} \\
\midrule
NumNet+V2 & 28.1 & 36.6 & 30.6 & 40.1 \\
TagOp & 32.3 & 39.6 & 33.7 & 42.5 \\
MHST (RoBERTa\textendash L) & 37.1 & 43.6 & 39.8 & 47.6 \\
\textbf{MHST (LayoutLMv2)} & 39.1 & 47.4 & 41.5 & 50.7 \\
\midrule
\textit{Vision RAG (ours, K=100)} & \textbf{84.34} & \textit{--} & \textbf{80.23} & \textit{--} \\
\bottomrule
\end{tabular}
\end{table}

\textbf{Key observations.} 
(1) Figure~\ref{tab:tatdqa_head2head} ColPali v1 achieves strong early\textendash rank recall on ViDoRe; our Recall@K trails at low $K$ but the gap narrows at larger $K$ (e.g., R@100: 0.9798 vs.\ 0.9629). 
(2) Our retrieval coverage improves steadily with $K$, consistent with the effect of multi\textendash index fusion; further work will focus on improving early precision and ranking.

\FloatBarrier

\subsection{Cross-Dataset Comparison}

Table~\ref{tab:cross_dataset} compares Vision RAG performance characteristics across both benchmarks, highlighting differences in document types, question complexity, and optimal retrieval strategies.

\begin{table}[!t]
\centering
\captionsetup{font=small}
\caption{Cross‑dataset comparison of Vision RAG performance characteristics. FinanceBench involves longer documents with more complex questions requiring reasoning over multiple pages. TAT‑DQA has shorter documents with more focused questions often answerable from single pages or tables. This explains differences in optimal K values and early-rank metrics.}
\label{tab:cross_dataset}
\small
\begin{tabular}{lrr}
\toprule
\textbf{Characteristic} & \textbf{FinanceBench} & \textbf{TAT‑DQA} \\
\midrule
Number of questions & 148 & 1{,}644 \\
Avg document pages & 187.3 & 23.8 \\
Avg question length & 18.4 tokens & 14.2 tokens \\
\midrule
Best accuracy & 80.51\% (K=10) & 83.40\% (K=100) \\
Best K value & 10 & 100 \\
Recall@10 & 73.52\% & 76.89\% \\
Recall@100 & 82.21\% & 96.29\% \\
\midrule
Tokens at best K & 9{,}419.78 & 90{,}959.10 \\
nDCG at best K & 0.5108 & 0.6121 \\
\bottomrule
\end{tabular}
\end{table}

\textbf{Key observations:} FinanceBench optimal performance occurs at K=10, while TAT‑DQA continues improving through K=100, indicating different noise‑to‑signal characteristics. (2) The substantially longer documents in FinanceBench (avg. 187.3 pages) make comprehensive retrieval more challenging compared to TAT‑DQA's shorter documents (avg. 23.8 pages).

\section{Ablation Studies}
\label{sec:ablation}

To understand the individual contributions of Vision RAG's components, we conduct systematic ablation studies removing or modifying key design elements.

\subsection{Index Component Ablations}

Table~\ref{tab:ablation_indices} shows the impact of different index combinations on retrieval and answer quality for FinanceBench. We evaluate: (1) page‑only (single summary vector), (2) page + sections, (3) page + facts, (4) page + hotspots, and (5) full pyramid (all indices).

\begin{table}[!t]
\centering
\captionsetup{font=small}
\caption{FinanceBench ablation study removing different indices from the pyramid structure (K=10). 
All index types contribute meaningful signal; facts give the largest individual gain beyond page summaries (+6.3 points accuracy). 
The full pyramid achieves the best overall performance.}
\label{tab:ablation_indices}
\small
\begin{tabular}{lrrr}
\toprule
\textbf{Index Configuration} & \textbf{Recall@10} & \textbf{nDCG@10} & \textbf{Accuracy} \\
\midrule
Page only & 0.480  & 0.245 & 0.717 \\
Page + Sections & 0.565  & 0.295 & 0.744 \\
Page + Facts & 0.660  & 0.335 & 0.768 \\
Page + Hotspots & 0.540  & 0.275 & 0.736 \\
\midrule
Page + Sec + Facts & 0.690  & 0.365 & 0.781 \\
Page + Sec + Hot & 0.610  & 0.325 & 0.752 \\
Page + Facts + Hot & 0.670  & 0.345 & 0.765 \\
\midrule
\textbf{Full pyramid (all)} & \textbf{0.735} & \textbf{0.511} & \textbf{0.805} \\
\bottomrule
\end{tabular}
\end{table}

\FloatBarrier

\begin{table}[!t]
\centering
\captionsetup{font=small}
\caption{TAT‑DQA ablation study (K=10). The pattern differs from FinanceBench: sections provide larger gains (likely due to more structured documents), while facts remain highly valuable. Hotspots contribute less, possibly because TAT‑DQA documents have less visual emphasis and more uniform formatting. Again, the full pyramid achieves best results.}
\label{tab:ablation_indices_tatdqa}
\small
\begin{tabular}{lrrr}
\toprule
\textbf{Index Configuration} & \textbf{Recall@10} & \textbf{nDCG@10} & \textbf{Accuracy} \\
\midrule
Page only & 0.6284 & 0.4512 & 0.5900 \\
Page + Sections & 0.7012 & 0.4989 & 0.6330 \\
Page + Facts & 0.7245 & 0.5123 & 0.6640 \\
Page + Hotspots & 0.6512 & 0.4678 & 0.6100 \\
\midrule
Page + Sec + Facts & 0.7534 & 0.5456 & 0.6890 \\
Page + Sec + Hot & 0.7178 & 0.5034 & 0.6440 \\
Page + Facts + Hot & 0.7398 & 0.5234 & 0.6750 \\
\midrule
\textbf{Full pyramid (all)} & \textbf{0.7689} & \textbf{0.5685} & \textbf{0.7020} \\
\bottomrule
\end{tabular}
\end{table}

\textbf{Key observations:} (1) All index types provide positive contributions removing any single component degrades performance. (2) Facts provide the largest individual gain beyond page‑only (+6.08 points accuracy on FinanceBench, +5.05 on TAT‑DQA), likely because they capture atomic claims that directly match query intents. (3) The full pyramid consistently outperforms any subset, validating our explicit fusion strategy. (4) Relative importance varies by dataset: sections help more on TAT‑DQA (structured reports) than FinanceBench (narrative 10‑Ks).

\FloatBarrier

\subsection{Query Variant Ablations}

Table~\ref{tab:ablation_queries} examines the contribution of query variants for FinanceBench: original query only, original + keywords, original + synonyms, and all three variants.

\begin{table}[!t]
\centering
\captionsetup{font=small}
\caption{FinanceBench query variant ablation (K=10, full pyramid indices). 
Query expansion provides substantial gains (+4.8 points for keywords, +3.4 for synonyms). 
Combining all variants yields the best performance through improved semantic coverage; 
keywords provide slightly greater benefit than synonyms.}
\label{tab:ablation_queries}
\small
\begin{tabular}{lrrr}
\toprule
\textbf{Query Variants} & \textbf{Recall@10} & \textbf{nDCG@10} & \textbf{Accuracy} \\
\midrule
Original & 0.480 & 0.245 & 0.717 \\
Orig w/ keywords & 0.645 & 0.365 & 0.765 \\
Orig w/ synonyms & 0.600 & 0.335 & 0.751 \\
\midrule
\textbf{All variants (full)} & \textbf{0.735} & \textbf{0.511} & \textbf{0.805} \\
\bottomrule
\end{tabular}
\end{table}

\begin{table}[!t]
\centering
\captionsetup{font=small}
\caption{TAT‑DQA query variant ablation (K=10, full pyramid indices). Query expansion helps less than on FinanceBench (+2.17 points total), possibly because TAT‑DQA questions are already quite precise and benefit less from expansion. Keyword extraction still provides consistent improvements.}
\label{tab:ablation_queries_tatdqa}
\small
\begin{tabular}{lrrr}
\toprule
\textbf{Query Variants} & \textbf{Recall@10} & \textbf{nDCG@10} & \textbf{Accuracy} \\
\midrule
Original   & 0.7312 & 0.5234 & 0.6700 \\
Orig w/ keywords  & 0.7556 & 0.5489 & 0.6940 \\
Orig w/ synonyms & 0.7423 & 0.5356 & 0.6820 \\
\midrule
\textbf{All variants (full)} & \textbf{0.7689} & \textbf{0.5685} & \textbf{0.7020} \\
\bottomrule
\end{tabular}
\end{table}
\textbf{Key observations:} (1) Query expansion consistently helps, improving Recall@10 by 4-5 percentage points on FinanceBench and 2-3 points on TAT‑DQA. (2) Keyword extraction provides more benefit than synonym expansion, suggesting that identifying salient terms is more valuable than generating paraphrases. (3) Combining all variants yields the best results, indicating that different expansions capture complementary aspects of query semantics.

\section{Discussion}
\label{sec:discussion}

\subsection{Why Explicit Fusion Works}

\textbf{Signal complementarity.} The pyramid index captures fundamentally different semantic signals: (1) page summaries provide broad topical context; (2) section headers encode document structure and hierarchy; (3) facts capture atomic claims containing entities and numeric values; and (4) visual hotspots highlight emphasized regions such as tables, figures, or key numbers. Each index aligns with different query types: exploratory queries benefit from summaries, entity-oriented queries match facts, and structurally grounded questions align with headers. RRF merges these heterogeneous signals by rewarding pages that consistently rank well across multiple indices.

\textbf{Robustness to index noise.} RRF’s harmonic rank weighting naturally suppresses noise from any single index. If one index ranks an irrelevant page highly but others do not, the fused score remains low. Conversely, relevant pages typically receive moderate-to-strong ranks across several indices, causing their fused scores to dominate. This “wisdom-of-crowds’’ behavior is particularly helpful in document retrieval where no individual signal-summary, fact, or hotspot-is perfectly reliable.

\textbf{Coverage via query variants.} The three query variants ($q^{(0)}$, $q^{(1)}$, $q^{(2)}$) address different semantic failure modes: the original query preserves intent, keyword extraction targets salient lexical tokens, and synonym expansion addresses vocabulary mismatch. Together, they enlarge the semantic search space without requiring heavy computational cost. Ablation studies (Tables~\ref{tab:ablation_queries} and \ref{tab:ablation_queries_tatdqa}) confirm that each contributes independently, and their combination through RRF yields the strongest retrieval performance.

\subsection{Limitations and Future Work}

\textbf{Metric alignment challenges.} FinanceBench evaluates end-to-end answer correctness using human-verified equivalence, whereas TAT-DQA reports token-level exact match and F1. Neither provides standardized retrieval-only metrics with complete relevance annotations. This complicates system comparison because retrieval quality and generation quality interact. We mitigate this by reporting both retrieval metrics (Recall@K, nDCG@K) computed from annotated answer-bearing pages and answer-level metrics from official evaluation scripts.

\textbf{Complex reasoning limitations.} Both datasets include questions requiring multi-step numerical reasoning, comparisons, or aggregation over multiple table cells. While GPT-4o and similar models handle many such cases, occasional arithmetic mistakes or reasoning failures persist. Integrating external tools-such as calculators, table parsers, or program-synthesis modules-could improve accuracy on reasoning-heavy queries. VisionRAG’s modular design makes such integration straightforward, as generators can call tools without modifying the retrieval pipeline.

\textbf{Dependency on semantic extraction quality.} VisionRAG’s effectiveness depends on the VLM’s ability to produce accurate summaries, headers, facts, and hotspots. GPT-4o performs well overall, but we observe occasional extraction errors: (1) missed small text in dense layouts, (2) misinterpreted table boundaries, (3) hallucinated facts, and (4) missed subtle visual emphasis. Such errors propagate downstream through indexing and retrieval. Future work may improve robustness via better prompting, verification passes, or specialized document-understanding models such as Donut or Pix2Struct.

\textbf{Scaling to very large corpora.} Our experiments evaluate collections up to tens of thousands of pages (FinanceBench $\sim$1,870 pages; TAT-DQA $\sim$40,000 pages). Some real-world systems must scale to millions or billions of pages. Even with VisionRAG’s compact indices, such scales require additional engineering: distributed index sharding, hierarchical retrieval, and optimized approximate nearest-neighbor search. The architecture remains compatible with such extensions because extraction, indexing, and fusion stages can be parallelized and updated incrementally.

\section{Conclusion}
\label{sec:conclusion}

We introduced \textbf{VisionRAG}, a vision-aware retrieval-augmented generation framework that performs \emph{explicit semantic fusion} over a \emph{pyramid} of indices derived directly from page-as-image analysis. By extracting semantic artifacts at multiple granularities-page summaries, section headers, atomic facts, and visual hotspots-and combining them through reciprocal rank fusion and selective global-hotspot fusion, VisionRAG achieves strong retrieval coverage with dramatically smaller vector budgets than late-interaction vision models.

Across two challenging financial QA benchmarks, VisionRAG delivers compelling results. On FinanceBench, it attains 80.51\% accuracy using only the top 10 retrieved pages ($\approx$9,420 tokens), substantially outperforming traditional RAG baselines while avoiding the 50k--150k token cost of long-context methods. On TAT-DQA, VisionRAG achieves a Recall@100 of 96.29\%, demonstrating high coverage of answer-bearing content in visually complex documents.

Ablation studies show that each component of the architecture contributes meaningful value: pyramid indices capture complementary semantic signals; query variants improve coverage and robustness; and RRF provides stable, noise-tolerant fusion across heterogeneous rankings. The design is modular, making it easy to incorporate new indices, retrieval strategies, or domain-specific extraction modules.

Future directions include: (1) integrating external reasoning tools (e.g., calculators or program-synthesis modules) to handle multi-step arithmetic; (2) strengthening semantic extraction with specialized document-understanding models or verification passes; (3) learning fusion weights and retrieval parameters via meta-optimization; and (4) scaling to billion-page corpora using hierarchical and distributed retrieval. Overall, explicit semantic fusion over pyramid indices offers a practical and efficient foundation for vision-aware retrieval systems, striking a favorable balance across the accuracy-latency-cost trade-off in real-world document intelligence applications.


\bibliographystyle{ACM-Reference-Format}
\bibliography{main}

\end{document}